\documentclass[11pt,preprint]{aastex}
\usepackage{graphicx}
\usepackage{enumerate}
\usepackage{amsmath}

\def\bfnabla{{\mbox{\boldmath $\nabla$}}}

\renewcommand\bv{{\mbox{\boldmath $v$}}}

\newcommand\bP{{\mbox{\boldmath $P$}}}
\newcommand\bT{{\mbox{\boldmath $T_g$}}}

\newcommand\bFg{{\mbox{\boldmath $F_g$}}}

\newcommand\bI{{\mbox{\boldmath $I$}}}
\renewcommand{\b}[1]{\boldsymbol{#1}}
\def\<{\,\langle\langle}
\def\>{\,\rangle\rangle}

\newcommand{\ba}{\begin{eqnarray}}
\newcommand{\ea}{\end{eqnarray}}

\slugcomment{August 8th, 2012}

\shortauthors{Y.-F. Jiang, M. A. Belyaev, J. Goodman \& J. Stone}

\author{Yan-Fei Jiang\altaffilmark{1,\ast}, Mikhail Belyaev\altaffilmark{1}, Jeremy Goodman\altaffilmark{1} \& James M. Stone\altaffilmark{1}}
\affil{$^1$Department of Astrophysical Sciences, Princeton
University, Princeton, NJ 08544, USA} 
\affil{\text{$\ast$Corresponding author. Email address: yanfei@astro.princeton.edu}}

\begin{document}

\title{A New Way to Conserve Total Energy for Eulerian Hydrodynamic
  Simulations with Self-Gravity}

\begin{abstract}
We propose a new method to conserve the total energy to round-off error 
in grid-based codes for hydrodynamic simulations with self-gravity. A formula for the
energy flux due to the work done by the the self-gravitational
force is given, so the change in total energy can be written in conservative form. 
Numerical experiments with the code Athena show that the total energy
is indeed conserved with our new 
algorithm and the new algorithm is second order accurate. 
We have performed a set of tests that show the numerical 
errors in the traditional, non-conservative algorithm can affect the
dynamics of the system. 
The new algorithm only requires one extra solution of the Poisson 
equation, as compared to the traditional algorithm which includes
self-gravity as a source term. If the Poisson solver takes a negligible fraction of the total 
simulation time, such as when FFTs are used, the new algorithm 
is almost as efficient as the original method. This new algorithm
is useful in Eulerian hydrodynamic
simulations with self-gravity, especially when results are sensitive 
to small energy errors, as for radiation pressure dominated flow.
\end{abstract}


\section{Introduction}
Self-gravity plays an important role in many astrophysical flows. 
For instance, it is central to
massive star formation \citep[e.g.,][]{MckeeOstriker2007,Krumholzetal2009}, supernova
explosions \citep[e.g.,][]{Nordhausetal2010}, planet formation 
in protoplanetary disks \citep[e.g.,][]{Armitage2011}, and structure formation in the early universe 
\citep[e.g.,][]{Bertschinger1998}. 
Numerical modeling of these systems requires 
that self-gravity be implemented correctly in hydrodynamical
simulations, especially in problems for which energy balance is important.
For example, the fate of a collapsing cloud depends on 
the total energy which is the sum of the potential, internal, and
kinetic energies. For some systems, e.g. those having a relativistic
equation of state with adiabatic gamma, $\gamma \approx 4/3$, the
total energy can be much smaller than either the potential or internal
energies in hydrostatic equilibrium. Thus, a small
numerical error in the
computation of the potential energy can lead to a large
fractional error in the total energy, potentially causing a bound system to become
unbound \citep[e.g.,][]{JiangGoodman2011} or vice versa. 
Energy conservation in a grid-based code with self-gravity 
is usually not guaranteed. 
The goal of this paper is to propose a new 
algorithm for grid-based codes, which conserves the sum of the internal, kinetic, and
potential energies to round-off error, allowing accurate
simulations of self-gravitating systems to be performed. 

The change in the momentum and the energy due to self-gravity are
typically added to the equations of hydrodynamics as source terms. In
this case, the Euler equations read
\begin{eqnarray}
\label{rhoeq}
\frac{\partial\rho}{\partial t}+\bfnabla\cdot(\rho \bv)&=&0,\\
\label{peq}
\frac{\partial( \rho\bv)}{\partial t}+\bfnabla\cdot({\rho \bv\bv+\bP})
&=&-\rho\bfnabla\phi, \ \\
\label{eeq}
\frac{\partial{E}}{\partial
  t}+\bfnabla\cdot\left[(E+P)\bv)\right]&=&-\rho\bv\cdot\bfnabla\phi,
\ \\
\label{poisson}
\bfnabla^2\phi=4\pi G\rho.
\end{eqnarray}
Here $\rho,\ \bv, P$ are density, velocity, and pressure
respectively, and $E$
is the sum of the internal and kinetic energies. $G$ is the gravitational 
constant, and $\phi$ is the gravitational potential, which is related
to the density via the Poisson 
equation (\ref{poisson}). We can also define the total energy as
\ba
E_{tot} \equiv E + \frac{1}{2} \rho \phi.
\ea
$E_{tot}$ is a {\it globally} conserved quantity so that
\ba
\label{globalcons}
\frac{\partial}{\partial t} \int d^3{\bf x} E_{tot} = 0
\ea
for any isolated self-gravitating system. Expression
(\ref{globalcons}) can be derived by integrating equation (\ref{eeq})
over all of space and applying the divergence theorem (see e.g., \citealt{BinneyTremaine2008}, section 2.1).

Equations (\ref{peq}) and (\ref{eeq}) are written in 
non-conservative form. Therefore, momentum and energy 
are usually not conserved to roundoff error when they are solved numerically. 
It is well known that if momentum is not conserved for 
a self-gravitating system, then the shape of the object cannot be kept very well 
when it moves across the simulation box \citep[e.g.,][]{Edgareal2005,Taskeretal2008}.
In order to conserve momentum, it is now common practice to write 
the momentum equation in conservative form as
\begin{eqnarray}
\frac{\partial( \rho\bv)}{\partial t}+\bfnabla\cdot({\rho \bv\bv+\bP+\bT}) =0,
\end{eqnarray}
where the gravitational tensor $\bT$ is
\begin{eqnarray}
\bT=\frac{1}{4\pi G}\left[\bfnabla\phi \bfnabla \phi-\frac{1}{2}(\bfnabla\phi)\cdot(\bfnabla\phi)\bI\right],
\label{equation_momentum}
\end{eqnarray} 
and $\bI$ is the identity tensor. This equation is mathematically equivalent to the 
original momentum equation (\ref{peq}) with use of equation (\ref{poisson}). However 
equation (\ref{equation_momentum}) can be solved in a completely
different way numerically. Momentum 
change due to gravitational acceleration is no longer included as a source term. Instead, it is 
included as a flux term via $\bT$, which means the total momentum can
be conserved to round-off error numerically. This has already been implemented in 
many grid-based codes, such as Athena.

The improvement we propose in this paper is to also write the energy 
equation with self-gravity (\ref{eeq}) in conservative form. Using
this form of the equation, we devise a numerical 
algorithm that conserves the total energy with self-gravity to round-off error.

\section{Formula for the energy flux}
We begin by rewriting the energy equation (\ref{eeq}) in
the following form
\begin{eqnarray}
\frac{\partial}{\partial t}\left({E+\frac{1}{2}\rho\phi}\right)+\bfnabla\cdot\left[(E+P)\bv+\bFg)\right]=0. 
\label{equation_energy}
\end{eqnarray}
As long as the flux due to self-gravity $\bFg$ falls off faster than
$r^{-2}$ at large distances and the system is spatially bounded,
equation (\ref{equation_energy}) automatically
implies that equation (\ref{globalcons}) is satisfied. We now solve
for the form of $\bFg$. 

Using equation (\ref{eeq}), equation (\ref{equation_energy}) can be written as
\begin{eqnarray}
\frac{\partial}{\partial t}\left(\frac{1}{2}\rho\phi\right)+\bfnabla\cdot \bFg=\rho\bv\cdot\bfnabla\phi.
 \end{eqnarray}
 Using the continuity equation, this becomes
 \begin{eqnarray}
\label{eqmid}
\bfnabla\cdot\bFg=\frac{1}{2}\bfnabla\cdot\left(\rho\bv\right)\phi+\rho\bv\cdot\bfnabla\phi-\frac{1}{2}\rho\dot{\phi}.
\label{dFg}
 \end{eqnarray}
 Here $\dot{\phi}\equiv\partial\phi/\partial t$ is the time rate of change
 of the potential. Differentiating the Poisson equation with
 respect to time yields
 \begin{eqnarray}
\label{ddtpoiss}
 \bfnabla^2\dot{\phi}=4\pi G\dot{\rho}=-4\pi G\bfnabla\cdot\left(\rho\bv\right).
 \label{dotphi}
 \end{eqnarray}
Substituting this in equation (\ref{eqmid}) gives
 \begin{eqnarray}
\label{momflux}
\bfnabla\cdot\bFg&=&\bfnabla\cdot\left(\rho\bv\right)\phi+\rho\bv\cdot\bfnabla\phi
+\frac{1}{8\pi G}\left(\phi\bfnabla^2\dot{\phi}-\dot{\phi}\bfnabla^2\phi\right)\\ \nonumber
&=&\bfnabla\cdot\left[\rho\bv\phi+\frac{1}{8\pi G}\left(\phi\bfnabla\dot{\phi}-\dot{\phi}\bfnabla\phi\right)\right],
 \end{eqnarray}
so one form of the energy flux due to self-gravity is
\begin{eqnarray}
\bFg=\frac{1}{8\pi G}\left(\phi\bfnabla\dot{\phi}-\dot{\phi}\bfnabla\phi\right)+\rho\bv \phi.
\end{eqnarray}
Note that this form for the energy flux is not unique, and as discussed in 
Appendix \ref{alternative}, alternative forms are available, which 
also satisfy equation (\ref{momflux}). However, only the divergence of the gravitational 
tensor and energy flux are used to evolve the system. The gravitational tensor 
and energy flux themselves are not used directly. Thus, as long as equation 
(\ref{momflux}) is satisfied, different forms of the energy flux are
mathematically equivalent. We note that a similar formula
for energy flux due to self-gravity has been proposed by \cite{Pen1998}.

In order to calculate the energy flux, we need to solve
equation (\ref{ddtpoiss}) for $\dot{\phi}$.  Note that this equation
has the same form as the Poisson
equation, which means it can be solved with the same numerical technique
(such as FFT for periodic boundary conditions).  Details on how to calculate the energy 
flux $\bFg$ are given in the next section.

\section{Numerical Implementation}
We implement the equations of hydrodynamics with self-gravity in
conservative form using Athena \citep[e.g.,][]{Stoneetal2008}, which is a MHD code that uses an
unsplit, higher order Godunov scheme. Since the implementation of
self-gravity is unchanged by the addition of a magnetic field, we
focus on the hydrodynamic case here. With self-gravity, 
the following steps are needed in addition to the original 
algorithm described in \cite{Stoneetal2008}.

First, at time step $n$ we need to compute the gravitational 
potential $\phi^n$ by solving the Poisson equation using the density
distribution.  For periodic boundary conditions, 
this can be done efficiently using the Fast Fourier Transform (FFT). 

Second, after we get the left and right states at the cell interface for 
each direction (step 1 of 6.1 in \citealt{Stoneetal2008}), we need to add the 
change due to self-gravity to the left and right states at the half time step $\delta t/2$. 
If primitive variables (density, velocity and pressure) are used for the reconstruction, 
we only need to add the gravitational acceleration $-\bfnabla\phi$ to the left and 
right velocity. No gravitational source terms need to be added to the
pressure. 

Third, after we get the fluxes at the interfaces from the Riemann solvers (step 7 in \citealt{Stoneetal2008}),
 we solve equation (\ref{dotphi}) for $\dot{\phi}^{n+1/2}$, and 
 the same numerical technique as for solving the Poisson equation can
 be used. We just need to replace $\rho$ on the right hand side of the
 Poisson equation with 
 $-\bfnabla\cdot\left(\rho\bv\right)$, where $\rho\bv$ is the density
 flux from the Riemann 
 solvers at each interface.  The $\dot{\phi}^{n+1/2}$ we get is at the
cell centers and is the time-averaged value for time step $n$. 
 
 Fourth, we update the density from time step $n$ to $n+1$ and calculate the new potential 
 $\phi^{n+1}$ based on the updated density distribution. Then, we use the averaged potential 
 $(\phi^n+\phi^{n+1})/2$ and $\dot{\phi}^{n+1/2}$ to
 calculate the gravitational tensor $\bT$ and the energy flux
 $\bFg$. The cell-centered potential $\phi$ is spatially-averaged to get the potential at cell interfaces.
 
 Finally, we update the momentum and the total energy with $\bT$ and $\bFg$. 
 At each cell $(i,j,k)$, if the cell size is $\delta x\times\delta y\times\delta z$, 
then the momentum change (take $\rho v_x$ for example) due to self-gravity is  
 \begin{eqnarray}
 ( \rho v_x)_{i,j,k}^{n+1} =(\rho v_x)_{i,j,k}^{n} + 
&-& \frac{\delta t}{\delta x} \left( {\bT_{xx}}^{n+1/2}_{i+1/2,j,k}
 - {\bT_{xx}}^{n+1/2}_{i-1/2,j,k} \right) \nonumber \\
&-& \frac{\delta t}{\delta y} \left( {\bT_{yx}}^{n+1/2}_{i,j+1/2,k}
 -{\bT_{yx}}^{n+1/2}_{i,j-1/2,k} \right) \nonumber \\
&-& \frac{\delta t}{\delta z} \left( {\bT_{zx}}^{n+1/2}_{i,j,k+1/2}
 - {\bT_{zx}}^{n+1/2}_{i,j,k-1/2} \right).
\end{eqnarray}
and the change in the total energy is
 \begin{eqnarray}
  E_{i,j,k}^{n+1} = E_{i,j,k}^{n} + \left[\frac{1}{2}\left(\rho\phi\right)^{n}_{i,j,k} -\frac{1}{2}\left(\rho\phi\right)^{n+1}_{i,j,k}\right]
&-& \frac{\delta t}{\delta x} \left( {\bFg_x}^{n+1/2}_{i+1/2,j,k}
 - {\bFg_x}^{n+1/2}_{i-1/2,j,k} \right) \nonumber \\
&-& \frac{\delta t}{\delta y} \left( {\bFg_y}^{n+1/2}_{i,j+1/2,k}
 - {\bFg_y}^{n+1/2}_{i,j-1/2,k} \right) \nonumber \\
&-& \frac{\delta t}{\delta z} \left( {\bFg_z}^{n+1/2}_{i,j,k+1/2}
 - {\bFg_z}^{n+1/2}_{i,j,k-1/2} \right).
\end{eqnarray}
Note that the definition of $E$ does not include potential energy.
  
\section{Numerical Tests}
To show that our algorithm is stable, second order accurate, and
conserves the total energy to machine precision, we perform several tests in one, two, and
three dimensions. We also compare the results of the new algorithm
with a non-conservative one that has self-gravity
added as a source term. In all cases, the non-conservative algorithm use the conservative 
form of the momentum --- but not energy --- equation. 
In the non-conservative algorithm, the gravitational potential $\phi^{n}$ is first 
calculated based on density at time step $n$.  
After the density is updated, the new potential $\phi^{n+1}$ 
at time step $n+1$ is calculated. 
Then the energy source term 
is added as $-\left(\rho\bv\right)^{n+1/2}\cdot\bfnabla\left(\phi^{n}+\phi^{n+1}\right)/2$, where 
$\left(\rho\bv\right)^{n+1/2}$ is the average of the left and right
sides of the density flux. 
In this way, the non-conservative algorithm can achieve second order accurate 
as long as energy error due to self-gravity is negligible. 

\subsection{Jeans Collapse in 1D and 2D}

All of our tests in 1D and 2D are based on the Jeans instability problem, where self-gravity
is the dominant force for the dynamics. We
initialize the background state to be a self-gravitating, uniform
density medium and assume the fluid is adiabatic with
sound speed $c_s^2 = \gamma P/\rho$. On top of
this background state, we initialize a small-amplitude ($\delta
\rho/\rho = 10^{-6}$) normal mode perturbation with
wavevector $k$. The amplitude of the perturbation changes with time as $
e^{i\omega t}$, where $\omega$ is given by the Jeans dispersion
relation
\ba
\omega^2 = k^2c_s^2 - 4\pi G \rho.
\label{jeanseq}
\ea
The Jeans length is given by $\lambda_J = \sqrt{\pi c_s^2/G\rho}$,
so that a normal mode perturbation with wavelength $\lambda <
\lambda_J$ results in a propagating wave,
whereas one with $\lambda > \lambda_J$ yields an instability that
saturates by forming dense clumps. We focus on the unstable case
for comparing the conservative and non-conservative algorithms.
Table \ref{simtable} lists the details of each
of our simulations and shows the relevant numerical parameters used.


\begin{table}[tp]
\centering
\begin{tabular}{ccccccccccc}
\hline
\hline
sim & dim & alg & $N$ & $\lambda$ & $\lambda_J$ & $M_0$ & $\gamma$  \\
\hline
A & 1 & c  & 256 & 256 & 128 & 0 & 5/3  \\
B & 1 & c  & 16 & 16 & 32 & 0 & 5/3 \\
B2 & 2 & c & 23x23 & 16 & 32 & 0 & 5/3 \\
C & 1 & c  & 32 & 32 & 64 & 0 & 5/3 \\
C2 & 2 & c & 45x45 & 32 & 64 & 0 & 5/3 \\
D & 1 & c  & 64 & 64 & 128 & 0 & 5/3  \\
D2 & 2 & c & 91x91 & 64 & 128 & 0 & 5/3 \\
E & 1 & c  & 128 & 128 & 256 & 0 & 5/3 \\
E2 & 2 & c & 181x181 & 128 & 256 & 0 & 5/3 \\
F & 1 & c  & 256 & 256 & 512 & 0 & 5/3 \\
F2 & 2 & c & 362x362 & 256 & 512 & 0 & 5/3 \\
G & 1 & c & 512 & 512 & 1024 & 0 & 5/3 \\
H & 2 & nc & 22x22 & 16 & 8 & 0 & 5/3 \\
I & 2 & nc  & 45x45 & 32 & 16 & 0 & 5/3 \\
J & 2 & nc & 91x91 & 64 & 32 & 0 & 5/3\\
Jc & 2 & c & 91x91 & 64 & 32 & 0 & 5/3  \\
K & 2 & nc & 181x181 & 128 & 64 & 0 & 5/3 \\
L & 2 & nc & 362x362 & 256 & 128 & 0 & 5/3\\
M & 2 & nc & 91x91 & 64 & 32 & 0 & 4/3 \\
N & 2 & nc& 181x181 & 128 & 64 & 0 & 4/3 \\
O & 2 & nc & 362x362 & 256 & 128 & 0 & 4/3 \\
P & 2 & c  & 91x91 & 64 & 32 & 10 & 5/3 \\
\hline
\end{tabular}
\caption{From left to right, the columns designate the simulation label, the number of
dimensions, the algorithm used (c for conservative, nc for
nonconservative), the number of cells ($N_x \times N_y \times N_z$), the
wavelength of the initial perturbation measured in cells (all cells have the same size), 
the Jeans wavelength measured in cells, the Mach number of
the background fluid, and the adiabatic index of the gas.  The Courant
number for all simulations is $0.8$, and the
HLLC Riemann solver is used for all simulations.} 
\label{simtable}
\end{table}

\subsubsection{1D tests}
\begin{figure}[!h]
  \centering
  \vspace{-6cm}
\includegraphics[width=1.1\textwidth]{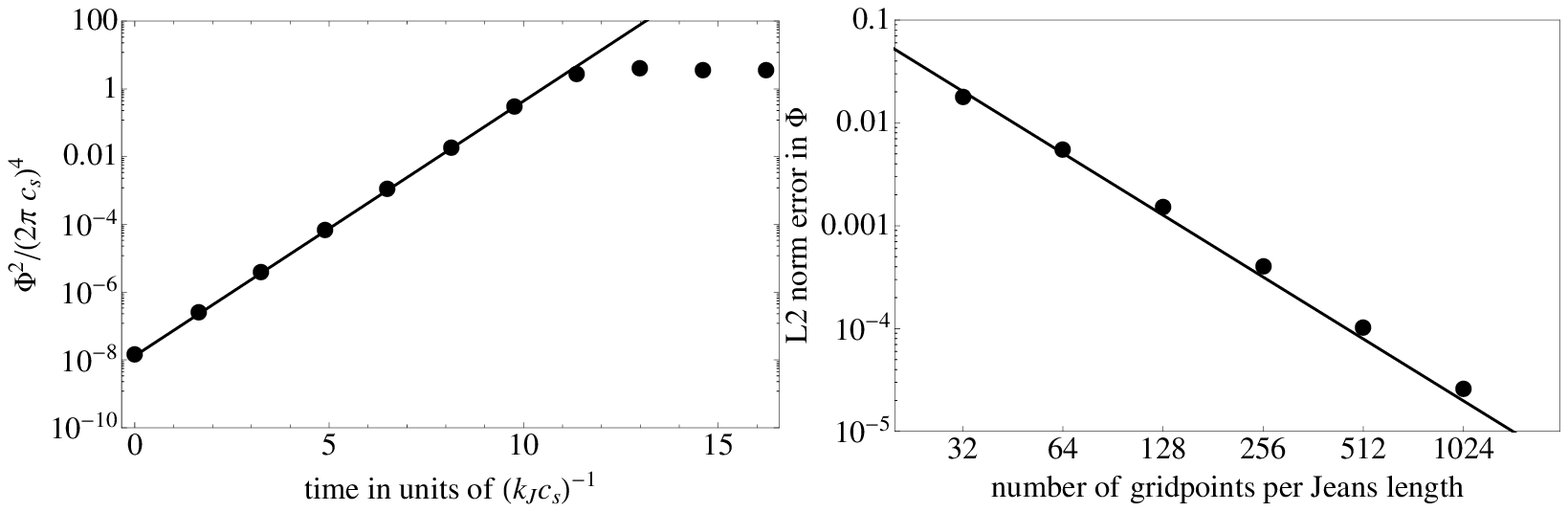}
\vspace{-9cm}
  \caption{\emph{Left}: Comparison of the analytical and numerical solutions for
  simulation A. The analytical solution for the linear stage of growth is shown by the line, and the
  numerical solution is shown
  with points. At late times, the numerical solution deviates from the
  linear analytical solution because the instability becomes non-linear
  and saturates.
  \emph{Right}: Convergence study for the conservative algorithm. The points
  correspond to simulations B --- F, and a line with slope indicating
  second order convergence is plotted for reference.} 
  \label{phifig}
\end{figure}

The first tests we perform are an accuracy and a convergence test in
1D. In Figure \ref{phifig}, the left panel shows the time-evolution of the square of the
perturbed gravitational potential integrated over the simulation
domain for simulation A, which is for an unstable mode
having $\lambda=2\lambda_J$. 
The points show values from the simulations
with time measured in units of $(k_J c_s)^{-1}$ and distance
measured in units of $\lambda_J$. The gravitational potential is thus
measured in units of $(\lambda_J k_J c_s)^2 = (2 \pi c_s)^2$. The solid line 
shows the analytical solution obtained from the dispersion relation
(\ref{jeanseq}). It is clear that the code accurately
captures the exponential growth in the linear phase of the instability
through to saturation. The right panel of Figure  \ref{phifig} 
shows convergence of the new algorithm with
resolution. The initial perturbation has $\lambda = \lambda_J/2$ and
corresponds to a
stable perturbation. The points are for simulations B-F (the only
thing that varies between the simulations is resolution) and the x-axis
shows the number of cells per Jeans length.
The y-axis shows the error in the $\mathcal{L}_2$ norm between the
numerical and analytical solutions for $\Phi$ after one period ($t = 2
\pi/\omega$). The error in the $\mathcal{L}_2$ norm can be represented as
\ba
\epsilon(\mathcal{L}_2) = \frac{\sqrt{\sum_i\left(\Phi_i -
    \tilde{\Phi}_i\right)^2}}{\sqrt{\sum_i \Phi_i^2}}, 
\ea 
where $\Phi_i$ is the value of the exact solution for the potential at point
$i$, $\tilde{\Phi}_i$ is the value of the numerical solution, and
the summation runs over all points in the simulation domain.
For reference, we also plot a
line whose slope indicates second order convergence, and which 
confirms that the new algorithm does indeed converge at second order.


To test the conservation properties of the new algorithm in
comparison with the old one, we compute the fractional energy error at
timestep $n$ to be $|E^n-E^0|/E^0$, where
\ba 
E^n = \sum_i{\left(\frac{1}{2} \rho_i^n (v_i^n)^2 + \frac{P_i^n}{\gamma - 1} +
  \frac{1}{2} \rho_i^n \Phi_i^n\right)}.
\ea
Indeed, our new algorithm conserves the total 
energy to round-off error. However, in 1D, the 
original algorithm also conserves the total energy 
to very high precision.  
Thus, we need to
go to a higher number of dimensions to demonstrate the superior energy
conservation properties of the new algorithm.

\subsubsection{2D linear wave tests}
To see how our new algorithm and the 
non-conservative algorithm behave in 2D, 
we again initialize an
unstable eigenmode with $\lambda = 2 \lambda_J$, just as in the 1D
case. Now, however, we choose $k_x/k_y =
1$ so that the wavefronts are not
aligned with the grid. 

We perform two kinds of tests for the 2D case, which are analogous to
the tests in the 1D case. The first is a convergence test of the
conservative and non-conservative algorithms, for a stable mode ($\lambda < \lambda_J$). The
second, is a test of the energy conservation for the two algorithms for an unstable mode ($\lambda >
\lambda_J$). 

\begin{figure}[!h]
  \centering
\includegraphics[width=.8\textwidth]{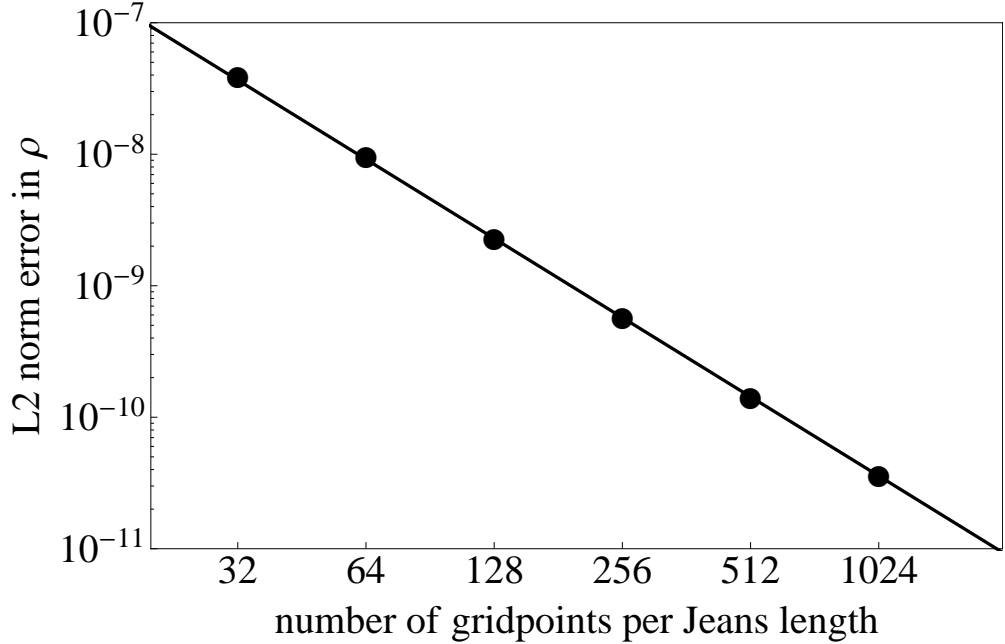}
  \caption{ Convergence study for the conservative algorithm in 2D. The points
  correspond to simulations B2 --- F2, and a line with slope indicating
  second order convergence is plotted for reference. The non-conservative algorithm 
  gives almost the same result for this stable propagation linear mode.} 
  \label{2Dconverge}
\end{figure}

Figure \ref{2Dconverge} shows the results of the convergence test for
the conservative algorithm (simulations B2-F2). On a log-log graph, we
plot the $\mathcal{L}_2$ norm error in $\rho$ as a function of
resolution after one oscillation
period for a mode having $\lambda = \lambda_J/2$, and an initial
amplitude of $\delta \rho/ \rho = 10^{-6}$. The points denote the
simulation results, and the slope of the solid line shows second order
convergence. It is clear that the conservative algorithm does indeed
converge at second order. We have tested that the
non-conservative algorithm also converges at second order for this the stable mode.
Because the results are essentially identical to the conservative algorithm, we
do not plot them. 

We now describe tests of energy conservation, using an unstable normal
mode. In 2D, in order to make the normal mode collapse to a
filament as opposed to a sheet, we increase the density perturbation
in the center of the simulation box by 2\% relative to the edges. 
We find that the 
energy conservation properties of the non-conservative
algorithm are drastically degraded relative to the 1D case. Table
\ref{errortable} shows the fractional energy error for the
nonconservative 2D, $\gamma = 5/3$
simulations (H --- L), and the nonconservative 2D, $\gamma = 4/3$
simulations (M --- O). It is clear from the table, that in 2D the energy
error for the
non-conservative algorithm can be of order the total energy,
although the error does decrease with increasing
resolution. On the other hand, the conservative algorithm still
conserves the total energy to machine precision for the same set of
tests. 

Note that the initial total energy
(ignoring the energy of the perturbation) is roughly $E =
P/(\gamma-1)$. Thus, the initial total energy for the $\gamma =4/3$
case is twice that for the $\gamma = 5/3$ case. This means that
although the fractional energy error for the $\gamma = 5/3$ case is
roughly double that of the $\gamma = 4/3$ case, the absolute value of
the accumulated energy error is roughly the same. We have also made
sure that the total energy is conserved for the conservative algorithm
if we advect the background medium with respect to the grid at
Mach 10 (simulation P).

The main cause of the discrepancy between the 1D and the 2D case
is likely to be twofold. First, the wavevector is no longer
aligned with the grid, so the 2D case is less
symmetric than the 1D case. Second, the collapse proceeds further in
the 2D case: to a filament in 2D as opposed to a sheet in 1D. 
As self-gravitational energy is a global quantity, 
the energy error caused by the source terms of self-gravity has 
different dependence on time step and grid size compared with 
the normal local truncation error. This energy error will be at 
a maximum when the gravitational potential changes most rapidly. 
When the total numerical error is dominated by the gravitational 
energy error for the non-conservative algorithm, 
Table \ref{errortable} shows that the error scales roughly as $\mathcal{O}(1/N)$. 
However, when the energy error from self-gravitational source terms is 
much smaller than the usual truncation error as in the case of Figure \ref{2Dconverge}, 
the non-conservative algorithm shows second order convergence with resolution.

\begin{table}[t]
\centering
\begin{tabular}{cc|cc}
\hline
Label & Error & Label & Error \\
\hline
H & $.65$ & - & - \\
I & $.32$ & - & - \\
J & $.18$ & M & $.09$ \\
K & $.09$ & N & $.05$ \\
L & $.05$ & O & $.03$ \\
\hline
\end{tabular}
\caption{The energy error for the non-conservative simulations in 2D 
as a function of resolution (increasing resolution going
  down). The error shown is the fractional error in the total energy
  between the beginning of the simulation and when the perturbation
  has virialized. The simulations H, I, J, K, L are done with 
  adiabatic index $\gamma=5/3$ while the simulations 
  M, N, O are done with adiabatic index $\gamma=4/3$.  
   Note that in the case $\gamma = 4/3$, the initial
  total energy is twice that of the case $\gamma = 5/3$ (neglecting
  the energy of the perturbation).} 
  \label{errortable}
\end{table}



\begin{figure}[!h]
  \centering
\includegraphics[width=.5\textwidth]{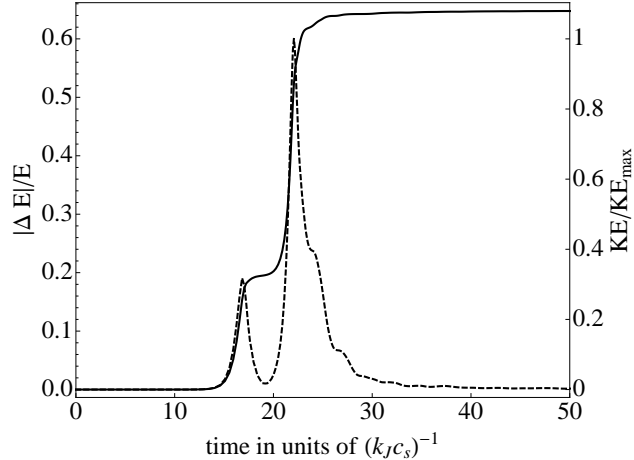}
  \caption{The solid line shows the fractional error in the total energy as a function of
  time for simulation H. The dashed line shows the kinetic energy as a fraction of
  the maximum kinetic energy in the simulation. It is clear that the
  error accumulates when the kinetic energy is highest. The plateau
  between $t=10$ and $t=20$ corresponds to the phase in which the collapsing
  object has already finished collapsing to a sheet, but is just starting to
  collapse to a filament.}
    \label{errortimeevol}
\end{figure}

Fig. \ref{errortimeevol} shows the time-evolution of the energy error
for simulation H. Overplotted is the kinetic energy in the
simulation, normalized to the maximum kinetic energy. It is evident that most of the the error is
accumulated in a short interval around the time when the kinetic energy, and
hence the infall velocity are large. This is also
when the gravitational potential changes most rapidly, making the
energy error large for the non-conservative algorithm.  

\begin{figure}[!h]
  \centering
  \vspace{-2cm}
  \includegraphics[width=0.9\textwidth]{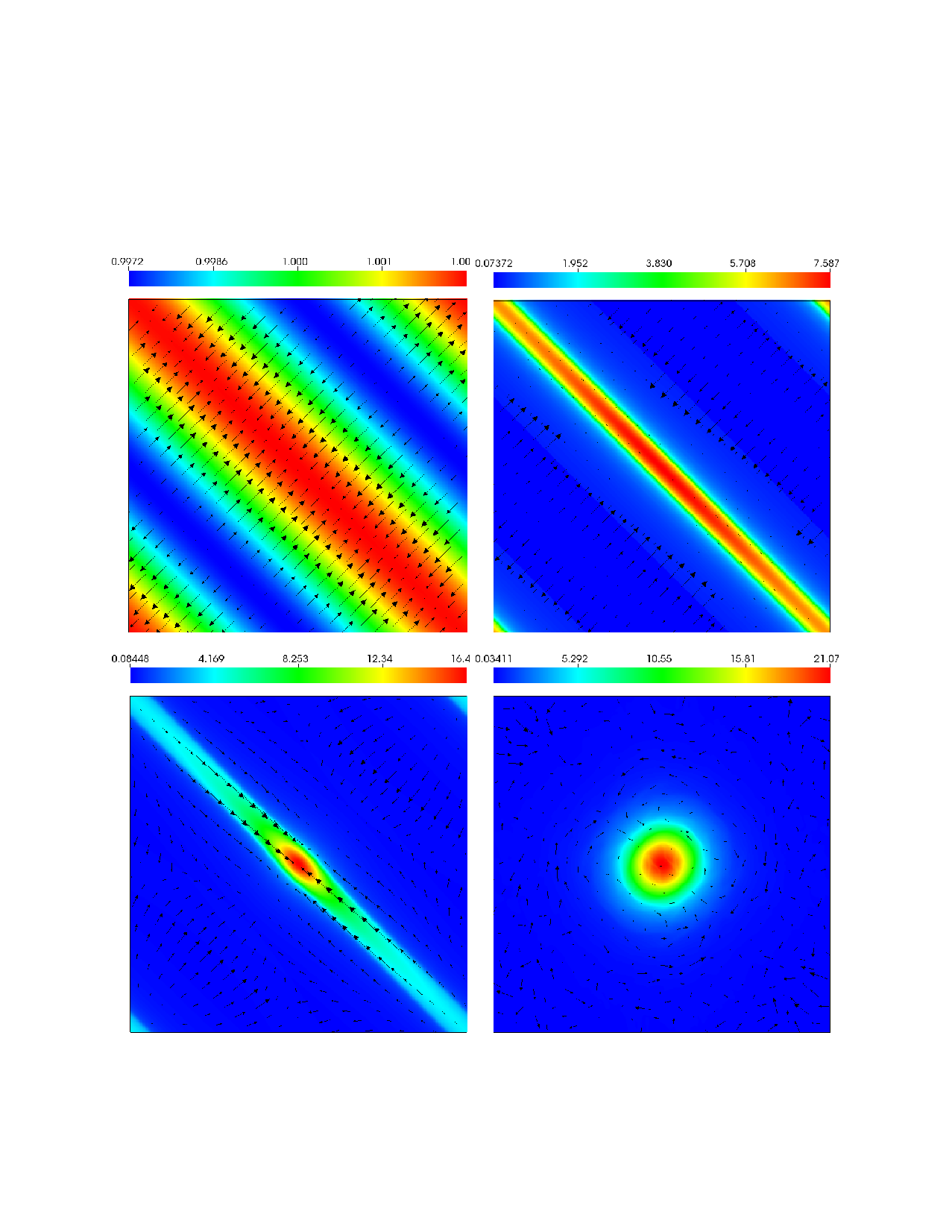}
  \vspace{-2cm}
  \caption{Snapshots of the density and the velocity field from
  simulation L. The scale of the density for each figure is
  given by the corresponding color bar, and the initial background
  density is $\rho = 1$. The magnitude scale of the velocity vectors is arbitrary,
  and varies from figure to figure. The upper left panel is taken at
  $t = 9.2 \ (k_Jc_s)^{-1}$, during the linear stage of the collapse. The upper right
  panel shows the simulation at $t = 18 \ (k_J c_s)^{-1}$, in the non-linear regime when
  the density distribution has collapsed to a sheet. The lower left
  panel shows the collapse of the sheet to a filament at $t = 21 \ (k_J c_s)^{-1}$. The lower
  right panel shows the final, virialized filament at $t= 73 (k_Jc_s)^{-1}$.} 
  \label{collapseview}
\end{figure}

For reference, Fig. \ref{collapseview} shows three snapshots of the
density from the 2D simulation which has a resolution of 128 cells per
Jeans length. The
first snapshot is taken during the linear phase of exponential growth, the
second during the non-linear phase when matter is collapsing to a sheet,
the third during the phase when matter is collapsing to a filament, and
the fourth during the final, virialized state. 

\begin{figure}[!h]
  \centering
  \includegraphics[width=0.6\textwidth]{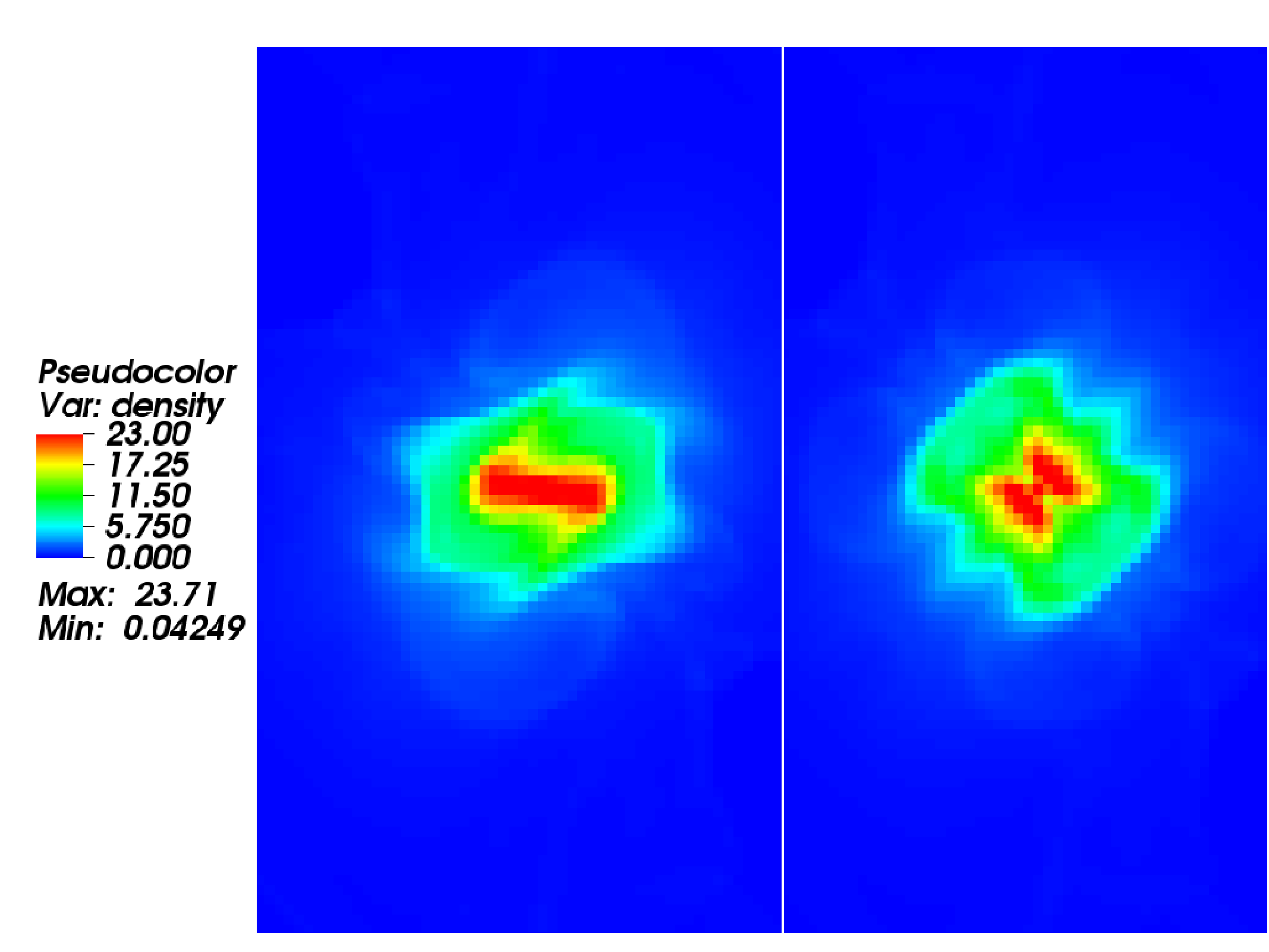}
  \caption{Comparison of the density distribution 
 for the conservative and non-conservative algorithm 
 for simulations J and Jc in Table \ref{simtable}. 
 The snapshots are taken at 
 time $t=1.1$, when the collapse stops and the energy error 
 reaches its maximum value for the simulation using the 
 non-conservative algorithm. The left panel is the density 
 distribution from the non-conservative algorithm while the 
 right panel is the result from the conservative algorithm. 
} 
  \label{densityprofiles}
\end{figure}

In order to see the possible effect on the dynamics due to the energy error, 
we compare the density distribution from simulations J and Jc. 
The two simulations have exactly the same parameters but simulation J
uses the non-conservative
algorithm, whereas simulation Jc uses the conservative one. Figure
\ref{densityprofiles} shows that
the density distributions are different depending on whether the
total energy is conserved or not. When the total energy is conserved
(right panel
of Figure \ref{densityprofiles}), the formed clump is symmetric, as it
should be, because it is formed
from a symmetric filament. However, for the non-conservative algorithm,
the symmetry
is lost (left panel of Figure \ref{densityprofiles}). Moreover,
rotation is generated
with the non-conservative algorithm, which shouldn't be the case,
since the initial angular momentum is zero. This is likely because the 
energy error of the non-conservative algorithm changes the distribution 
of gas pressure, which causes the rotation of the fluid.  

\subsection{Collapse of  a Polytropic Sphere in 3D}
The test that shows the difference between the conservative 
and non-conservative algorithms most clearly in 3D is the collapse
of a polytropic
sphere. This test is also relevant to many astrophysical systems where 
self-gravity is important, such as star formation and planet formation
in protoplanetary disks.
In 3D, the adiabatic index $\gamma=4/3$ is a critical case when the
total energy of a
hydrostatic self-gravitating sphere becomes zero. The dynamics of the
sphere will be sensitive to the 
energy error made in the numerical solution, which may change the sign
of the total energy in the solution. The $\gamma = 4/3$ case is
relevant for massive stars when  gravity is balanced by a radiation
pressure gradient \citep[e.g.,][]{JiangGoodman2011}.

We initialize a polytropic sphere with polytropic index $n=3$ in a
periodic domain, so
the radial density $\rho(r)$ and pressure $P(r)$ profiles are
given by
\begin{eqnarray}
\frac{1}{r^2}\frac{d}{dr}\left(\frac{r^2}{\rho}\frac{dP}{dr}\right)=-4\pi G\rho,\quad\quad \frac{P}{P_{c,0}}=\left(\frac{\rho}{\rho_{c,0}}\right)^{4/3}.
\end{eqnarray}
In our units, $4\pi G=1$, the central density is $\rho_{c,0}=1$, and
the central pressure in hydrostatic equilibrium is $P_{c,0}=8.09\times 10^{-5}$.
The adiabatic index $\gamma=1.36$ is chosen to be close to $4/3$ in
order to clearly demonstrate the
difference between the two algorithms. The background density in the
simulation is set to $0.005$. The diameter of the sphere 
is $0.16$, which is $20\%$ the size of the simulation box, and the
sphere is located at the center of the
simulation box. The resolution we use is $256^3$. 

Initially, we decrease the central pressure to be a fraction of the pressure in hydrostatic equilibrium and let
the sphere evolve under its own gravity. Let us estimate how much the size of the polytropic sphere should change 
due to a decrease in the central pressure.  Starting from the virial theorem and using simple scaling relations, it is straightforward to derive that in hydrostatic equilibrium
\begin{eqnarray}
\label{consteq}
M^{2-\gamma}R^{3\gamma-4} = \alpha(\gamma) P_{c,0}\rho_{c,0}^{-\gamma}.
\end{eqnarray}
Here $\alpha(\gamma)$ is a constant that depends only on $\gamma$, $R$ is the radius of the sphere, and $M$ is the total mass. If we assume homologous adiabatic collapse, then the right hand side of equation (\ref{consteq}) is a constant during the collapse. Suppose now that we reduce the initial pressure to be a fraction $\epsilon$ of the pressure in hydrostatic equilibrium. We can then solve for the final radius, $R$, in terms of the initial radius $R_0$ as
\begin{eqnarray}
\label{solveforradius}
\frac{R}{R_0} = \epsilon^\frac{1}{3\gamma - 4}
\end{eqnarray}
In our simulations, we use $\gamma=1.36$ and $\epsilon = .93$, which
gives $R/R_0 \approx .4$.
Thus, we expect to get a large-amplitude, spherically-symmetric, periodic solution. Moreover, since the fractional radius of the sphere is small compared to the box size, we expect spherical symmetry to be maintained quite well, even though we use periodic boundary conditions.

\begin{figure}[!h]
  \centering
\includegraphics[width=.4\textwidth]{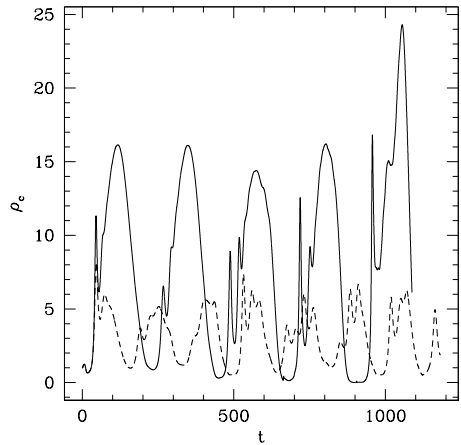}
  \caption{Time evolution of the central density in the polytropic
  sphere test from two simulations that use the same initial conditions
  but different numerical algorithms. The solid line is the result using the non-conservative
  algorithm whereas 
  the dashed line is the result using the conservative one. }
\label{rhoct}
\end{figure}

\begin{figure}[!h]
  \centering
 \vspace{-9cm}
 \hspace{-3cm}
\includegraphics[width=1.2\textwidth]{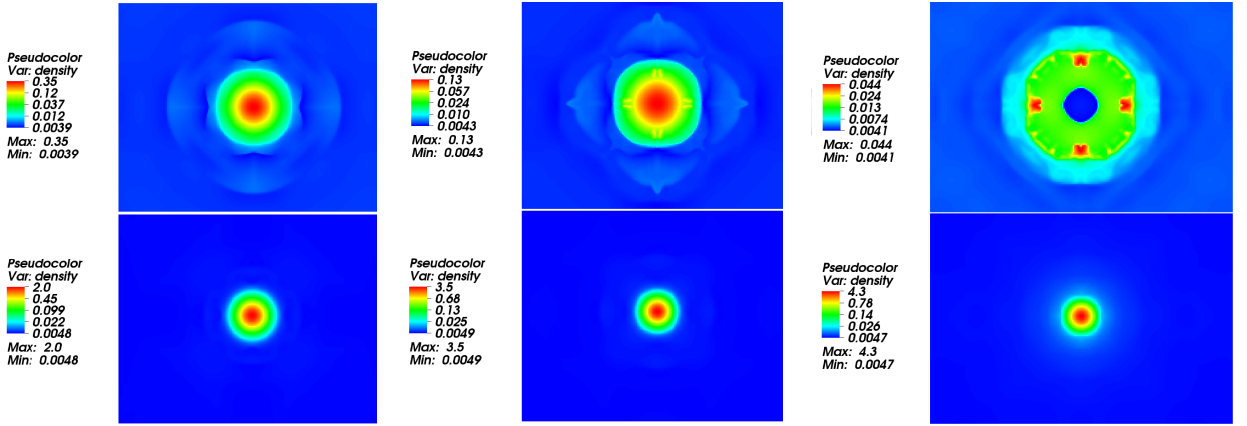}
\vspace{-9cm}
  \caption{Snaphsots of the density distributions from the polytropic
  sphere test taken at different times through the plane $y=0$ from two 
  simulations using different algorithms. From left
  to right, the times are $t=460, 680, 900$, and in each case, only the central region is shown.
  The two simulations start from exactly the same initial conditions. The top panels correspond
  to the
  non-conservative algorithm whereas the bottom panels correspond to the conservative one. 
  A central hole is created with the non-conservative algorithm and the
  simulation shows strong deviations from spherical symmetry.}
\label{3Ddensity}
\end{figure}

\begin{figure}[!h]
  \centering
\includegraphics[width=.4\textwidth]{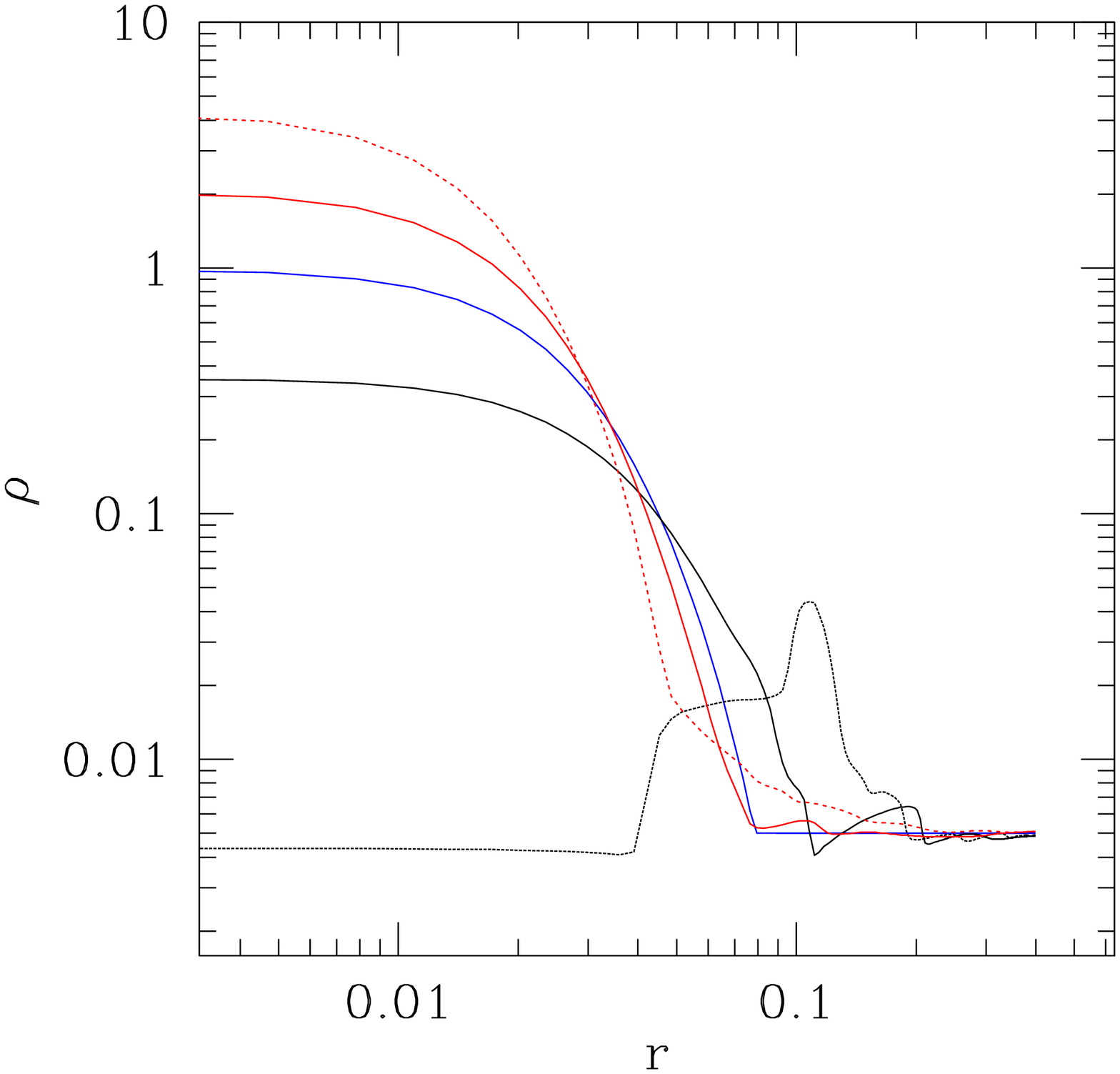}
  \caption{Density profiles at different times on the line $(x=0,y=0)$
  from the 3D polytropic sphere test using two different numerical
  algorithms. The blue line is the initial
  density profile, which is the same for both simulations. The black lines are results for 
  the simulation using the non-conservative algorithm whereas the red
  lines are results for the simulation using the 
  conservative algorithm. The solid black and red lines are measured
  at time $t=460$ while the dashed 
  black and red lines are measured at time $t=900$. }
\label{rhospatial}
\end{figure}

We start two simulations with exactly the same
initial conditions but
one uses the non-conservative algorithm whereas the other uses the
conservative one.
The time-evolution of the central density from the two 
simulations is shown in Figure \ref{rhoct}. Both simulations give
almost the same
result for $t<15$, but after the initial phase of the collapse, they show quite
different behaviors with the non-conservative algorithm attaining a
much larger peak central density. Snapshots of the density
distribution from the two simulations through the plane $y=0$ 
at times $t=460, t=680$, and $t=900$ are shown in Figure \ref{3Ddensity}. 
Radial profiles of the spheres at time $t=460$ and $t=900$ 
through the line $(x=0, y=0)$ are examined in Figure \ref{rhospatial}.  
At the early time, $t=460$, both algorithms are able to keep the spherical symmetry of the 
object, although non-spherical structures have already appeared
with the non-conservative algorithm (top panels of Figure \ref{3Ddensity}). These non-spherical
structures 
are amplified by time $t=680$. Eventually, the initial
density profile is destroyed for the non-conservative algorithm, 
and a low density hole is formed at the center of the sphere at time
$t=900$. 
On the other hand, the conservative algorithm (bottom panels of Figure \ref{3Ddensity}) 
yields a periodic solution and
keeps the spherical symmetry and initial profile of the 
sphere much better during the collapse than the nonconservative
one.  In conclusion, this test shows that for $\gamma \approx 4/3$ in 3D, the dynamics are
significantly affected by the choice of algorithm. The conservative algorithm is favored,
since it better preserves spherical symmetry throughout the collapse and yields a periodic solution 
as expected.

\section{Performance}
In order to calculate the time derivative of the potential, we need to 
solve one extra Poisson equation compared to the old method. If the
Poisson solver takes a significant fraction of
the total simulation time, the code will be slowed down by almost a factor of two 
for this new feature. If the time spent in the Poisson solver is
negligible, then
the new code is almost as efficient as the old one. For all our
simulations, we use periodic boundary conditions and solve the
Poisson equation using FFTs. In 
this case, the new algorithm is almost as efficient as the original method.

\section{Summary and Discussion}
We have developed a new algorithm to conserve total energy to round-off error 
for Eulerian hydrodynamical simulations incorporating self-gravity. 
We have implemented this new algorithm in Athena and shown that 
it conserves total energy to round-off error as expected. By comparing 
a set of tests in 1D, 2D and 3D with two different numerical algorithms 
for self-gravity (the new conservative algorithm and the traditional non-conservative algorithm), 
we have shown that the numerical error made in the traditional algorithm can 
change the dynamics significantly. From our numerical experiments, we
conclude that the
conservative algorithm will be important when a
small amount of energy error can
dramatically affect the dynamics, such as in radiation-dominated systems 
with an adiabatic index close to $4/3$. 

We tested our new algorithm by implementing it in Athena, which uses
an unsplit Godunov method to 
solve the equations of hydrodynamics. In principle, this 
algorithm can also be used in codes that implement an operator split
scheme, such as ZEUS \citep[][]{StoneNorman1992}.
In the application using ZEUS with self-gravity performed by 
\cite{JiangGoodman2011} , the energy error
using the traditional non-conservative algorithm was positive. 
If ZEUS is used to study problems of star formation with 
the non-conservative algorithm, gravitationally bound clumps can
acquire a positive total energy due to numerical errors and be
destroyed as a result of numerical heating. However, if our 
conservative algorithm is used, this will not happen, since the total
energy is conserved.

The new conservative algorithm is not necessary for all hydrodynamical
simulations incorporating self-gravity. 
We find that for cases when the energy error is not a concern, the
two algorithms give 
almost identical results. For example, for small
oscillations of a polytropic sphere with an adiabatic index of
$\gamma=5/3$, the two algorithms yield very similar temporal
and spatial behavior. In such cases, the old algorithm is just as
good as the new one and is potentially more efficient.

\section*{Acknowledgements}
This work started as an assignment for class APC524 in Princeton
University. We thank the anonymous referee for helpful 
comments to improve this paper. We also thank Ue-Li Pen and 
Adam Burrows for helpful discussions. 
Y.-F J. thanks Robert Lupton
and Clancy Rowley for helpful suggestions during the implementation of
this algorithm in Athena. JG was supported in part by the NSF Center for 
Magnetic Self-Organization under grant PHY-0821899.

\begin{appendix}
\section{Alternate forms}
\label{alternative}
If the only properties required of the gravitational energy flux are that it
satisfy equation (\ref{dFg}) and have the correct physical dimensions, then $\b{F}_g$ is not unique.
Energy conservation is not affected by $\b{F}_g\to\b{F}_g+\b{a}$ 
if $\b{\nabla\cdot a}=0$.  For example,
\begin{equation*}
  \b{a}= \frac{1}{4\pi G}\b{\nabla\times}\left(\phi\b{v\times\nabla}\phi\right).
\end{equation*}
More generally, for arbitrary vector fields $\b{b}$, the replacements
\begin{equation*}
  \left(\b{F}_g,\,E_g\right)\to \left(\b{F}_g +\partial_t\b{b},\ E_g-\b{\nabla\cdot b}\right)
\end{equation*}
preserve $\partial_t E_g + \b{\nabla\cdot F}_g$ and therefore preserve the
energy-conservation law.  In particular, if one chooses $\b{b}=-(16\pi G)^{-1}\b{\nabla}(\phi^2)$, then starting from equation 
(\ref{momflux}) for $\b{F}_g$ and
$\tfrac{1}{2}\rho\phi$ for $E_g$,  one obtains the alternate forms
\begin{equation}
  \label{eq:altforms}
  \b{F}_g = \rho\phi\b{v} - \frac{1}{4\pi G}\dot\phi\b{\nabla}\phi\,,\quad E_g= \rho\phi
  +\frac{1}{8\pi G}\left|\b{\nabla}\phi\right|^2\,.
\end{equation}
Because they differ by a divergence, the two forms of $E_g$ yield the same total
gravitational energy when integrated over all space.

Additional criteria are clearly needed to select the ``correct'' forms of $E_g$ and
$\b{F}_g$.   One such criterion is galilean covariance.
The energy density, energy flux, momentum density, and stress tensor of a nonselfgravitating fluid are
\begin{equation*}
  E=\frac{1}{2}\rho v^2 + u,\quad \b{F}=(E+p)\b{v},\quad \b{j}=\rho\b{v},
\quad\mathbf{T}=\rho\b{vv}+p\mathbf{I}\,,
\end{equation*}
where $u$ is internal energy per unit volume.
Under an infinitesimal galilean transformation $(\b{v},\rho,p,u)\to(\b{v}+\b{\delta V},\rho,p,u)$,
these densities and fluxes transform to first order in $\b{\delta V}$ as
\begin{align}
  \label{eq:Galtrans}
  E       &\to E + \b{j\cdot\delta V},&
   \b{j} & \to \b{j} + \rho\b{\delta V},\nonumber\\
 \b{F} &\to \b{F} + E\b{\delta V}+\mathbf{T}\b{\cdot\delta V}, &
\mathbf{T} &\to \mathbf{T} + 2\b{j\cdot\delta V}\,.
\end{align}
It seems reasonable to require that the gravitational
contributions to $E$, $\b{F}$, and $\b{T}$ should preserve the form of this
transformation.  Since the newtonian gravitational mass and momentum densities vanish
under any sensible definition, $E_g$ and $\mathbf{T}_g$ should be galilean invariants, while
\begin{equation}
  \label{eq:Galtrans_g}
\b{F}_g \to \b{F}_g + E_g\b{\delta V}+\mathbf{T}_g\b{\cdot}\b{\delta V}.
\end{equation}
Because $\b{\nabla}\phi$ and $\phi$ are galilean-invariant, $\mathbf{T}_g$ 
[equation (\ref{equation_momentum})] and
both forms of $E_g$ are invariant.  However, $\dot\phi\to\dot\phi-\b{\delta
  V\cdot\nabla}\phi$, from which one can show that the forms \eqref{eq:altforms} of $E_g$
and $\b{F}_g$ are compatible with \eqref{eq:Galtrans_g}, whereas the original forms
$E_g=\rho\phi/2$ and $\b{F}_g$ as given by equation (\ref{momflux}) are not.

No matter what definitions are used for $(E_g,\b{F}_g,\mathbf{T}_g)$, the same time
evolution results if the conservative equations are solved exactly, provided that the
definitions are mathematically equivalent to the original equations of motion (\ref{rhoeq}) --- (\ref{poisson}).
However, the spatial distribution of gravitational energy and stress will depend upon the
inertial frame used, and it is possible that numerical truncation errors may be sensitive
to this dependence.  The whole point of the present exercise is recast the equations so as
to improve energy conservation in finite-difference or finite-volume approximations.
Viewed in this light, numerical robustness becomes more important than formal elegance.
We mistrust the form \eqref{eq:altforms} of the gravitational energy density because it is
always non-negative in vacuum or near-vacuum regions: thus a small local error in the
estimate of $|\b{\nabla}\phi|^2/8\pi G$ could cause a large error in the update of the
kinetic or internal energies per unit mass.  The original form $E_g=\rho\phi/2$
does not have this defect because it tends to zero smoothly with the mass density.

If it is important both that the gravitational terms be locally galilean covariant and that
$E_g=\rho\phi/2$, then we may keep
equation (\ref{momflux}) for $\b{F}_g$  but replace the gravitational stress tensor (\ref{equation_momentum}) with
\begin{equation}
  \label{eq:altstress}
  \mathbf{\tilde T}_g= \frac{1}{8\pi G}\left[(\b{\nabla}\phi)(\b{\nabla}\phi)
- \phi \b{\nabla\nabla}\phi\right]+\frac{1}{2}\rho\phi\mathbf{I}\,.
\end{equation}
The difference between (\ref{equation_momentum}) and \eqref{eq:altstress} can be written as
\begin{equation}\label{eq:Tg_difference}
  \mathbf{\tilde T}_g - \mathbf{T}_g = \frac{1}{16\pi G}\left[\mathbf{I}\,\nabla^2 -\b{\nabla\nabla}\right]\phi^2.
\end{equation}
Since the divergence of \eqref{eq:Tg_difference} vanishes, the gravitational momentum
density remains zero if $\mathbf{T}_g$ is replaced by $\mathbf{\tilde T}_g$.

\end{appendix}

\bibliographystyle{apj}
\bibliography{self_gravity}

\begin{thebibliography}{12}
\expandafter\ifx\csname natexlab\endcsname\relax\def\natexlab#1{#1}\fi

\bibitem[{{Armitage}(2011)}]{Armitage2011}
{Armitage}, P.~J. 2011, \araa, 49, 195

\bibitem[{{Bertschinger}(1998)}]{Bertschinger1998}
{Bertschinger}, E. 1998, \araa, 36, 599

\bibitem[{{Binney} \& {Tremaine}(2008)}]{BinneyTremaine2008}
{Binney}, J., \& {Tremaine}, S. 2008, {Galactic Dynamics: Second Edition}, ed.
  {Binney, J.~\& Tremaine, S.} (Princeton University Press)

\bibitem[{{Edgar} {et~al.}(2005){Edgar}, {Gawryszczak}, \&
  {Walch}}]{Edgareal2005}
{Edgar}, R.~G., {Gawryszczak}, A., \& {Walch}, S. 2005, in Protostars and
  Planets V, 8005

\bibitem[{{Jiang} \& {Goodman}(2011)}]{JiangGoodman2011}
{Jiang}, Y.-F., \& {Goodman}, J. 2011, \apj, 730, 45

\bibitem[{{Krumholz} {et~al.}(2009){Krumholz}, {Klein}, {McKee}, {Offner}, \&
  {Cunningham}}]{Krumholzetal2009}
{Krumholz}, M.~R., {Klein}, R.~I., {McKee}, C.~F., {Offner}, S.~S.~R., \&
  {Cunningham}, A.~J. 2009, Science, 323, 754

\bibitem[{{McKee} \& {Ostriker}(2007)}]{MckeeOstriker2007}
{McKee}, C.~F., \& {Ostriker}, E.~C. 2007, \araa, 45, 565

\bibitem[{{Nordhaus} {et~al.}(2010){Nordhaus}, {Burrows}, {Almgren}, \&
  {Bell}}]{Nordhausetal2010}
{Nordhaus}, J., {Burrows}, A., {Almgren}, A., \& {Bell}, J. 2010, \apj, 720,
  694

\bibitem[{{Pen}(1998)}]{Pen1998}
{Pen}, U.-L. 1998, \apjs, 115, 19

\bibitem[{{Stone} {et~al.}(2008){Stone}, {Gardiner}, {Teuben}, {Hawley}, \&
  {Simon}}]{Stoneetal2008}
{Stone}, J.~M., {Gardiner}, T.~A., {Teuben}, P., {Hawley}, J.~F., \& {Simon},
  J.~B. 2008, \apjs, 178, 137

\bibitem[{{Stone} \& {Norman}(1992)}]{StoneNorman1992}
{Stone}, J.~M., \& {Norman}, M.~L. 1992, \apjs, 80, 753

\bibitem[{{Tasker} {et~al.}(2008){Tasker}, {Brunino}, {Mitchell}, {Michielsen},
  {Hopton}, {Pearce}, {Bryan}, \& {Theuns}}]{Taskeretal2008}
{Tasker}, E.~J., {Brunino}, R., {Mitchell}, N.~L., {Michielsen}, D., {Hopton},
  S., {Pearce}, F.~R., {Bryan}, G.~L., \& {Theuns}, T. 2008, \mnras, 390, 1267

\end{thebibliography}

\end{document}